\documentclass[aps,pra,epsfig,showpacs]{revtex4}
\usepackage{graphicx}

\def\be{\begin{equation}}
\def\ee{\end{equation}}
\def\bea{\begin{eqnarray}}
\def\eea{\end{eqnarray}}
\def\bma{\begin{mathletters}}
\def\ema{\end{mathletters}}
\def\bi{\begin{itemize}}
\def\ei{\end{itemize}}

\def\C{\hbox{$\mit I$\kern-.7em$\mit C$}}

\begin{document}

\title{Local Cloning of Entangled Qubits}

\author{Sujit K. Choudhary}
\email{sujit_r@isical.ac.in} \affiliation{Physics and Applied
Mathematics Unit, Indian statistical Institute, 203, B. T. Road,
Kolkata 700 108, India}

\author{Samir Kunkri}
\email{skunkri_r@isical.ac.in} \affiliation{Physics and Applied
Mathematics Unit, Indian statistical Institute, 203, B. T. Road,
Kolkata 700 108, India}

\author{Ramij Rahaman}
\email{ramij_r@isical.ac.in} \affiliation{Physics and Applied
Mathematics Unit, Indian statistical Institute, 203, B. T. Road,
Kolkata 700 108, India}

\author{Anirban Roy}
\email{anirb@qis.ucalgary.ca} \affiliation{17, Bhupen Bose Avenue,
Kolkata 700004, India}

\begin{abstract}
We discuss the exact cloning of orthogonal but entangled qubits
under local operations and classical communication. The amount of
entanglement necessary in blank copy is obtained for various
cases. Surprisingly this amount is more than $1$ ebit for certain
set of two nonmaximal but equally entangled states of two qubits
system. To clone any three two qubits Bell states at least $\log_2
3$ ebit is necessary.
\end{abstract}


\maketitle

\section{Introduction}
Classical states can always be cloned perfectly. But the quantum
no cloning theorem \cite{wooter} prohibits exact cloning of
nonorthogonal states. However, orthogonal quantum states can
always be cloned if one can perform a operation on the entire
system.

A common scenario in quantum information processing is where a
multipartite entangled state is distributed among a number of
spatially separated parties. Each of these parties are able to
perform only local operations on the subsystem they possess and
can send only classical information to each other. This is known
as LOCC (Local operation and classical communication). If we
restrict ourselves only to LOCC, further restrictions on cloning
apply. For example, the very obvious first restriction will be; an
entangled blank state is needed to clone an entangled state.
Moreover, entanglement of blank state should at least be equal to
the entanglement of the state to be cloned, or else entanglement
of the entire system will increase under LOCC which is impossible.
However, with a  sufficient supply of entanglement; entangled
states can be cloned by LOCC. For example, any arbitrary set of
orthogonal states of two qubits can be cloned with the help of $3$
ebit. Any set of two orthogonal states need only $2$ ebit.

The concept of entanglement cloning under LOCC was first
considered by Ghosh {\it et. al.} \cite{ghosh1} where it was shown
that for LOCC cloning of two orthogonal Bell states and four
orthogonal Bell states, 1-ebit and 2-ebit of entanglement is
neccessary and sufficient. Later many works have been done in this
direction \cite{anselmi,owari}, which involve maximally entangled
states. In this paper, we consider cloning of arbitrary but
equally entangled orthogonal states under LOCC and the following
interesting results
are found: \\

\noindent (i) $\log_2 3$ ebit in the blank copy is necessary to
clone any three Bell states.\\ \\
\noindent (ii) Local exact cloning of any two orthogonal entangled
states is not possible with the help of same entanglement unless
the states are maximally entangled.\\ \\
\noindent (iii) Even a maximally entangled state of two qubits may
not help as blank copy for cloning a set of two orthogonal
nonmaximal equally entangled states if these states lie in the
same plane.
\section{Cloning Bell states}
The four Bell states are given as:
\begin{equation}
\label{bells}
|B_{mn}\rangle=\frac{1}{\sqrt{2}}\sum_{j=0}^{1}e^{2\pi
ijn/2}|j\rangle|j\oplus m\rangle, n,m=0,1.
\end{equation}
where one qubit is held by Alice and the other is held by Bob.

In a very elegant way, Ghosh {\it et. al.} \cite{ghosh1} has shown
that any two Bell states can be cloned with the help of  $1$ ebit,
whereas to copy all the $4$ Bell states, one needs at least $2$
ebit of entanglement in the blank copy. Recently, Owari and
Hayashi \cite{owari} have shown that any three Bell states cannot
be cloned if one ebit free entanglement is supplied as resource.
We, in this section, by entanglement considerations, not only
prove the same but also provide the necessary
entanglement resource for such a cloning.\\
To obtain the necessary amount of entanglement needed in the blank
copy for local cloning (now and onwards by \textbf{`local
cloning'} or \textbf{`cloning'} we will mean \textbf{`exact
cloning under LOCC'}) of three Bell states, we will make use of
the fact that the relative entropy of entanglement cannot be
increased by any LOCC operation. The relative entropy of
entanglement for a bipartite quantum state $\rho$ is defined by
\cite{vedral}:
$$E_{R}(\rho)=\min_{\sigma\epsilon D(H)}~S(\rho \|\sigma)$$
Here D is the set of all separable states on the Hilbert space H
on which $\rho$ is defined and $S(\rho \|\sigma)$ (the relative
entropy of $\rho$ to $\sigma$) is given by $S(\rho\|\sigma)\equiv
tr(\rho \log_2 \rho)- tr(\rho \log_2 \sigma)$.\\
Let $\rho_1\in H^1$ and $\rho_2\in H^2$ be two quantum states and
let $E_{R}(\rho_1)=~S(\rho_1 \|\sigma_1)$,\\
 $E_{R}(\rho_2)=~S(\rho_2 \|\sigma_2)$; \emph{i.e.} $\sigma_1( \in
H_1 )$ and $\sigma_2( \in H_2 )$ are the two separable states
which minimize the relative entropies of $\rho_1$ and $\rho_2$
respectively. Let $\sigma$ be the separable state belonging to the
Hilbert space $H_1\otimes H_2$ which minimizes the relative
entropy of $\rho_1\otimes \rho_2 $. Then:\\
\begin{equation}
\label{rel1}
 E_{R}(\rho_1 \otimes \rho_2)\leq S(\rho_1 \otimes
\rho_2 \|\sigma_1 \otimes \sigma_2) \end{equation} equality holds
when
$\sigma_1 \otimes \sigma_2=\sigma$.\\
 It was known \cite{eisert} \\
\begin{equation}
\label{rel2}
 S(\rho_1 \otimes \rho_2 \|\sigma_1 \otimes
\sigma_2)=S(\rho_1 \|\sigma_1)+S(\rho_2 \|\sigma_2) \end{equation}
hence
\begin{equation}
\label{rel3} E_{R}(\rho_1 \otimes \rho_2)\leq S(\rho_1
\|\sigma_1)+S(\rho_2 \|\sigma_2)
\end{equation}
\emph{i.e.}
\begin{equation}
\label{rel4}
 E_{R}(\rho_1 \otimes \rho_2)\leq
E_R(\rho_1)+E_R(\rho_2)
\end{equation}
If cloning of three Bell states ({\it e.g.} $|B_{00}\rangle,
|B_{01}\rangle, |B_{10}\rangle$) is possible with a known
entangled state (say $|B\rangle$) as blank copy (resource), then
the following state
\\ $\frac{1}{3}[~|B_{00}^{\otimes 2}\rangle \langle
B_{00}^{\otimes 2}|~ + ~|B_{01}^{\otimes 2}\rangle \langle
B_{01}^{\otimes 2}|~+~|B_{10}^{\otimes 2}\rangle \langle
B_{10}^{\otimes 2}|~]$ along with the blank state $|B\rangle$
given as the input to the cloner will
provide the ouput as:\\
$$\rho_{in}\left(=\frac{1}{3}\left[~|B_{00}^{\otimes 2}\rangle \langle
B_{00}^{\otimes 2}|~ + ~|B_{01}^{\otimes 2}\rangle \langle
B_{01}^{\otimes 2}|~+~|B_{10}^{\otimes 2}\rangle \langle
B_{10}^{\otimes 2}|\right]\otimes |B\rangle \langle B|\right)$$
$$\longrightarrow~ \rho_{out}\left(= \frac{1}{3}[~|B_{00}^{\otimes
3}\rangle \langle B_{00}^{\otimes 3}|~ + ~|B_{01}^{\otimes
3}\rangle \langle B_{01}^{\otimes 3}|~+~|B_{10}^{\otimes 3}\rangle
\langle B_{10}^{\otimes 3}|~]\right)$$

We now compare the relative entropies of entanglement of
$\rho_{in}$ and $\rho_{out}.$ \\ From inequality (\ref{rel4}), we
have
$$E_R(\rho_{in})\leq E_R \left(\frac{1}{3}[~|B_{00}^{\otimes 2}\rangle \langle
B_{00}^{\otimes 2}|~ + ~|B_{01}^{\otimes 2}\rangle \langle
B_{01}^{\otimes 2}|~+~|B_{10}^{\otimes 2}\rangle \langle
B_{10}^{\otimes 2}|]\right)~+~E_R\left(|B\rangle \langle
B|\right)$$\\
As $E_R\left(\frac{1}{3}[~|B_{00}^{\otimes 2}\rangle \langle
B_{00}^{\otimes 2}|~ + ~|B_{01}^{\otimes 2}\rangle \langle
B_{01}^{\otimes 2}|~+~|B_{10}^{\otimes 2}\rangle \langle
B_{10}^{\otimes 2}|]\right)~\leq2-\log_{2}3$ \cite{ghosh}, hence:\\
$$E_R(\rho_{in})\leq 2-\log_{2}3+E_R\left(|B\rangle \langle
B|\right)$$
 At least 2 ebit of entanglement can be distilled from
$\rho_{out}$ \cite{foot1} and the distillable entanglement is
bounded above by $ E_R$, hence
$$E_R(\rho_{out})\geq 2.$$ But relative
entropy of entanglement cannot increase under LOCC, and in the
output we have at least $2$ ebit of relative entropy of
entanglement, hence, in order to make cloning possible, $\log_2 3$
ebit is necessary in the blank state. Any two qubit state (even a
two qubit maximally entangled state) cannot provide this necessary
amount of entanglement.

\section{Cloning arbitrary entangled states}
Any two equally entangled orthogonal states can lie either in same
plane:\\
(\textbf{I})
$$|\Psi_{1}\rangle=a|00\rangle~+~b|11\rangle$$$$
 |\Psi_{2}\rangle=b|00\rangle~-~a|11\rangle$$
\begin{center}
or in different planes:
\end{center}
(\textbf{II})$$|\Psi_{1}\rangle={a|00\rangle~+~b|11\rangle}$$
 $$|\Psi_{3}\rangle=a|01\rangle~+~b|10\rangle$$
 where a,b are real and unequal and $a^2+b^2=1$. \\

In both the cases, if one provide two entangled states, each
having same entanglement as in the original one, cloning will be
trivially possible. Here we investigate the nontrivial case when a
single entangled qubit state is supplied as blank copy.\\

\textbf{\underline{Case(I)} }\\ \\ Suppose there exists a cloning
machine which can clone $|\Psi_{1}\rangle$ and $|\Psi_{2}\rangle$
when a pure entangled qubit state
$|\Phi\rangle(=c|00\rangle+d|11\rangle;
 c^2+d^2=1$) is supplied to it as blank copy.
 Let us supply  an equal
 mixture of $|\Psi_{1}\rangle$ and $|\Psi_{2}\rangle$ together
 with the blank state $|\Phi\rangle$ to it; \emph{i.e.}. the state
 input to the cloner is:
\begin{equation}
\label{in}
 \rho_{in}=\left[\frac{1}{2}P(|\Psi_{1}\rangle)+
\frac{1}{2}P(|\Psi_{2}\rangle)\right]\otimes
P\left(|\Phi\rangle\right) \end{equation}
The output of the cloner:\\
\begin{equation}
\label{out} \rho_{out}
=\frac{1}{2}P\left[|\Psi_{1}\rangle\otimes|\Psi_{1}\rangle\right]+
\frac{1}{2}P\left[|\Psi_{2}\rangle\otimes|\Psi_{2}\rangle\right]
\end{equation}
For proving impossibility of such a cloner, we make use of the
fact that Negativity, of a bipartite quantum state $\rho$,
$\emph{N}(\rho)$ cannot increase under LOCC \cite{vidal}.
$\emph{N}(\rho)$ is
given by \cite{zycz}\\
\begin{equation}
\label{nega}
 \emph{N}(\rho)\equiv \|\rho^{T_{B}} \|-1
\end{equation}
 where $\rho^{T_{B}}$ is the partial transpose with respect to
system B and $\|...\|$ denotes the trace norm which is defined as,
\begin{equation}
\label{partial}
\|\rho^{T_{B}}\|=tr(\sqrt{\rho^{T_{B}^{\dagger}}\rho^{T_{B}}}~)
\end{equation}
The negativity of the input state $\rho_{in}$ is
$$\emph{N}(\rho_{in})=2cd~\leq~1\hspace{3.2cm}$$
whereas, the negativity of the output is
$$\emph{N}(\rho_{out})=4a^2b^2+4\sqrt{a^2b^2(a^2-b^2)^2}\hspace{3.2cm}$$
The above cloning will not be possible as long as,
\begin{equation}
\label{condi1}
 cd < 2a^2b^2+2\sqrt{a^2b^2(a^2-b^2)^2}
\end{equation}

The above inequality has some interesting features, but the most significant feature is:\\
`Even a maximally entangled state of two qubits cannot help as
blank copy for a large number of pairs of nonmaximally entangled
state belonging to this class'(see the graph below). Numerical
calculations show that this is the case for $0.230\leq a\leq0.973$
(except for $a=\frac{1}{\sqrt{2}})$. This is surprising as
recently Kay and Ericsson \cite{kay} have given a protocol by
which all the pairs of states lying in different planes
(\textbf{II}) can be cloned with the help of 1 free ebit.\\
Other
important features are: (a) For $a=b=c=d=\frac{1}{\sqrt{2}}$ the
above inequality becomes an equality. This is consistent with an
earlier finding
\cite{ghosh1}that two maximally entangled bipartite state can be cloned with 1 free ebit.\\
(b) Inequality (\ref{condi1}) holds even for $c=a \neq d=b$(see
the graph below). This in turn implies that same amount of
entanglement ( as in the state to be cloned ) cannot help as blank
copy, for any pair of nonmaximally entangled states.\\

\includegraphics[width=10.0cm]{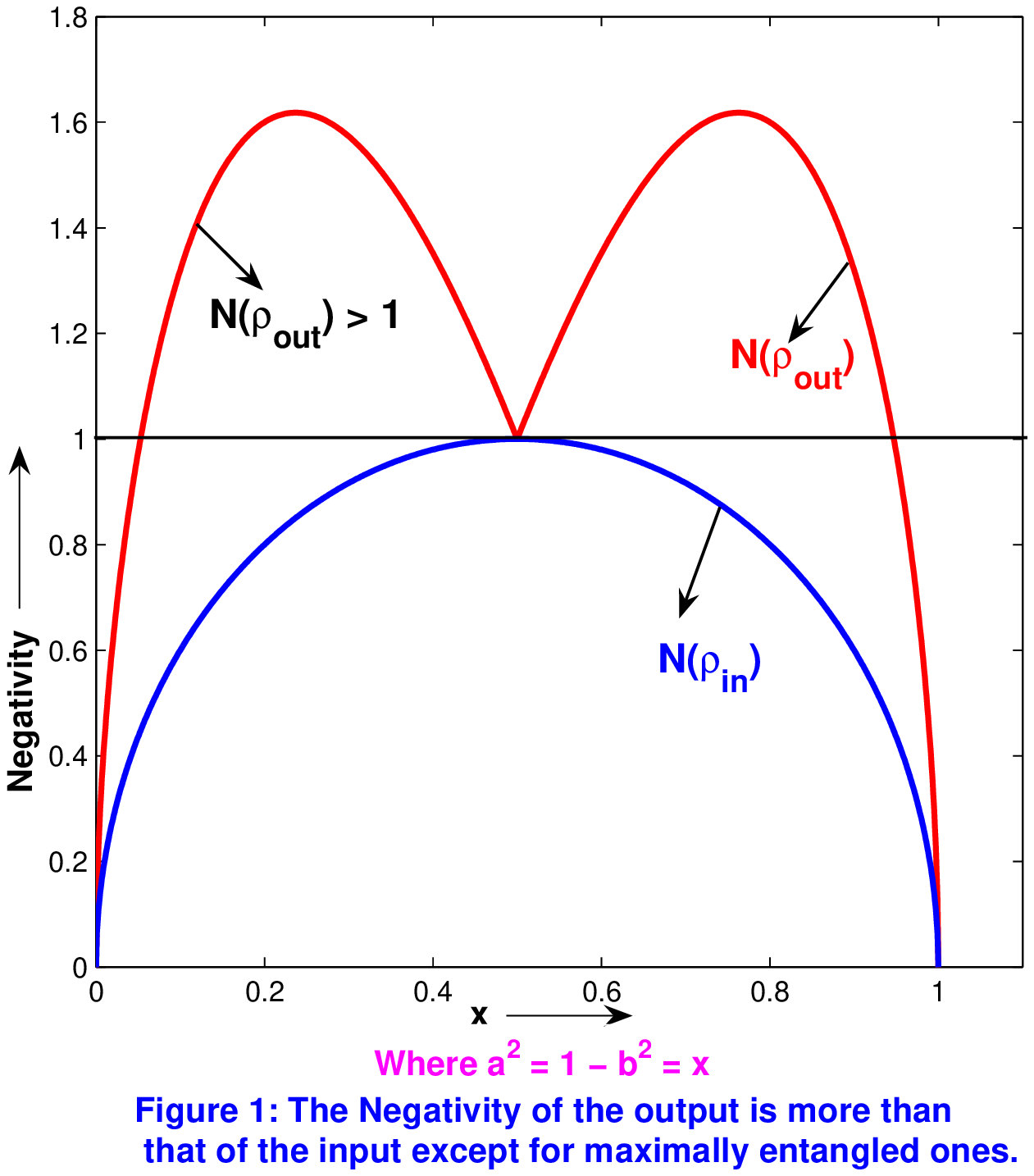}\\

\textbf{Case (II)}\\
 This time we suppose that our cloning machine
can clone $|\Psi_{1}\rangle$ and $|\Psi_{3}\rangle$ if a pure
entangled state $|\Phi\rangle(=c|00\rangle+d|11\rangle;
 c^2+d^2=1$) is used as blank copy.\\
Let the state supplied to this machine be:\\
$$\rho_{in}=\frac{1}{2}\left[P(|\Psi_{1}\rangle)~+~P(|\Psi_{3}\rangle)\right]\otimes P[|\Phi\rangle]$$ We then have
output of  the cloner as:\\
\\$$\rho_{out}=\frac{1}{2}P[|\Psi_{1}\rangle\otimes|\Psi_{1}\rangle]
+\frac{1}{2}P[|\Psi_{3}\rangle\otimes|\Psi_{3}\rangle]$$

Putting for $|\Psi_{1}\rangle$ $|\Psi_{3}\rangle$ and
$|\phi\rangle$ in the expression for $\rho_{in}$ and $\rho_{out}$
and making use of equations (\ref{nega}) and (\ref{partial}), we get: \\
$$\emph{N}(\rho_{in})=2cd \le 1 \hspace{3.2cm}$$
$$\emph{N}(\rho_{out})=2\sqrt{2(a^6b^2+a^2b^6)}\hspace{4.2cm}$$
From nonincrease  of negativity under LOCC it follows that as long
as
\begin{equation}
\label{condi2}
 cd < \sqrt{2(a^6b^2+a^2b^6)}
\end{equation}
the above cloning is not
possible.\\
(a) $a=b=c=d=\frac{1}{\sqrt{2}}$ turns this inequality into
an equality. This again is consistent with \cite{ghosh1}. \\
(b) If we put $c=a \neq d=b$ in the above inequality, \emph{i.e}
if we use same amount of entanglement (as in original states) then
too cloning remains
impossible as can be seen from the following graph:\\
\includegraphics[width=10.0cm]{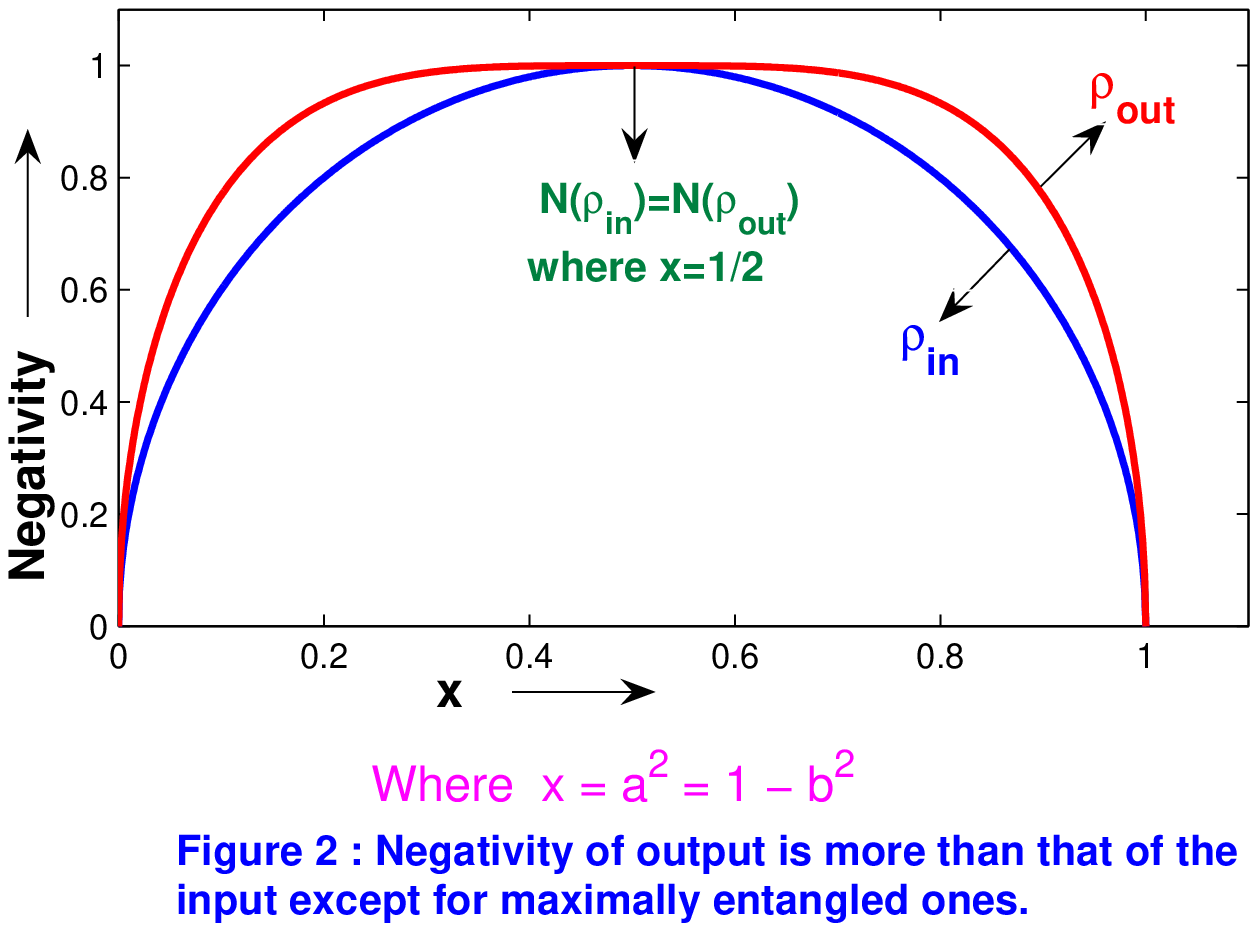}\\
(c) Here too the inequality (\ref{condi1}) shows that for any
entanglement in the original states, except the maximally ones,
the necessary entanglement in the blank copy is always higher. As
an example, for $a=\sqrt{0.3}$, ({\it i.e.} entanglement of the
state to be cloned $0.8813$), as long as $c < \sqrt{0.42}$, ({\it
i.e.} entanglement of blank copy $<0.9815$), cloning is not
possible.\\

\section{Conclusion}

In this paper we addressed the problem of LOCC cloning for
entangled states. To clone three Bell states, one need at least
$\log_2 3$ ebit in the blank state. So any two qubit state (pure
or mixed) cannot serve this purpose. We have also shown the blank
state needed should have more free entanglement than the original
ones, for cloning any pair of nonmaximal but equally entangled
orthogonal states. The necessary amount of entanglement in the
blanck state for such cloning to be possible is given by
inequalities (\ref{condi1}) and (\ref{condi2}). Interestingly this
necessary amount is more than 1 ebit for certain set of nonmaximal
but equally entangled states contrary to certain other sets for
which 1 ebit can serve as blank copy.

\begin{acknowledgments}
The authors acknowledge G.Kar for valuable suggestions. R.R
acknowledges the support by CSIR, Government of India, New Delhi.
\end{acknowledgments}


\begin{thebibliography}{99}

\bibitem{wooter} W. K. Wootters and W. H. Zurek, {\it Nature (London) } {\bf 229}, 802
(1982); D. Diekes, {\it Phys. Lett. } {\bf 92A}, 271 (1982); H. P.
Yuen,  {\it ibid. } {\bf 113A}, 405 (1986).

\bibitem{ghosh1} S. Ghosh, G. Kar and A. Roy, {\it Phys. Rev. A}
{\bf 69}, 052312 (2004).

\bibitem{anselmi} F. Anselmi, A. Chefles and M. Plenio, {\it New J.
Phys.} {\bf 6}, 164 (2004).

\bibitem{owari} M. Owari and M. Hayashi, {\it Phys. Rev. A} {\bf 74}, 032108 (2006).

\bibitem{vedral} V. Vedral and M. B. Plenio, {\it Phys. Rev. A} {\bf 57}, 1619
(1998); V. Vedral, M. B. Plenio, M. A. Rippin and P. L. Knight,
 {\it Phys. Rev. Lett. } {\bf 78}, 2275 (1997).

 \bibitem{eisert} J. Eisert, {\it eprint} quant-ph/0610253     and
 refernces therein.

\bibitem{ghosh} S. Ghosh, G. Kar, A. Roy, A. Sen(De) and U. Sen , {\it Phys.
Rev. Lett. } {\bf 87}, 277902 (2001).

\bibitem{foot1} The control-not operation $C$ is defined as
$C|i\rangle\otimes|j\rangle=|i\rangle\otimes|j\oplus i\rangle$,
and the bilateral control-not operation (BXOR) defined on
bipartite system as, ${\mathcal B}$ is ${\mathcal
B}|i\rangle_{A1}|r\rangle_{B1}\otimes|j\rangle_{A2}|s\rangle_{B2}=|i\rangle_{A1}|r\rangle_{B1}\otimes|j\oplus
i\rangle_{A2}|s\oplus r\rangle_{B2}.$ Denote ${\mathcal B(m,n)}$
as the BXOR operation performed on the mth pair (source) and the
nth pair (target), the following operation will give, ${\mathcal
B(1,3)}{\mathcal B(2,3)}|B_{mn}^{\otimes
3}\rangle=|B_{m,0}^{\otimes (2)}\rangle|B_{\oplus 3m,n}\rangle$
\cite{yang}. If this operation is applied on $\rho_{out}$, one get
$\frac{1}{3}[|B_{00}^{\otimes 2}\rangle \langle B_{00}^{\otimes
2}|\otimes [|B_{00}\rangle \langle B_{00}| + |B_{01}\rangle
\langle B_{01}|]+|B_{10}^{\otimes 2}\rangle \langle
B_{10}^{\otimes 2}|\otimes |B_{10}\rangle \langle B_{10}|]$. If
Alice and Bob do the measurement in $|0\rangle, |1\rangle$ basis
on the third copy and communicate, the results will be either
correlated or anticorrelated. When they they are correlated the
first two copies are in $|B_{00}\rangle$, and in other case they
are in state $|B_{10}\rangle$, therefore distilling two ebits in
this process.

\bibitem{yang} D. Yang and Y.-X. Chen, {\it Phys. Rev. A } {\bf 69}, 024302 (2004).



\bibitem{vidal} G. Vidal and R. F. Werner {\it Phys. Rev. A } {\bf 65}, 032314 (2002).

\bibitem{zycz} K. $\dot{Z}$yczkowski,  P. Horodecki, A. Sanpera, and M. Lewenstein, {\it Phys. Rev. A } {\bf 58}, 883 (1998).

\bibitem{kay} A. Kay and M. Ericsson, {\it Phys. Rev. A} {\bf 73},
012343 (2006).

\end{thebibliography}
\end{document}